\documentclass[10pt,journal,final,twocolumn]{IEEEtran}
\usepackage{amsmath,amsfonts}
\usepackage{algorithmic}
\usepackage{algorithm}
\usepackage{array}
\usepackage[caption=false,font=normalsize,labelfont=sf,textfont=sf]{subfig}
\usepackage{textcomp}
\usepackage{stfloats}
\usepackage{url}
\usepackage{verbatim}
\usepackage{graphicx}
\usepackage{cite}
\usepackage{hyperref}
\usepackage{booktabs} 
\usepackage{makecell} 
\usepackage{graphicx} 
\usepackage{multirow} 
\allowdisplaybreaks[4]
\hypersetup{hypertex=true,
            colorlinks=true,
            linkcolor=blue,
            anchorcolor=blue,
            citecolor=blue}
\makeatletter
\renewcommand{\maketag@@@}[1]{\hbox{\m@th\normalsize\normalfont#1}}
\makeatother

\def\BibTeX{{\rm B\kern-.05em{\sc i\kern-.025em b}\kern-.08em
    T\kern-.1667em\lower.7ex\hbox{E}\kern-.125emX}}

\begin{document}

\title{UAV Control and Communication Enabled Low-Altitude Economy: Challenges, Resilient Architecture and Co-design Strategies}

\author{Tianhao Liang, Nanchi Su, Yuqi Ping, Guangyu Lei, Xinglin Chen, Longyu Zhou, Tingting Zhang, Qinyu Zhang and Tony Q.S. Quek
\thanks{Tianhao Liang, Nanchi Su, Yuqi Ping, Guangyu Lei, Xinglin Chen, Tingting Zhang, and Qinyu Zhang are with the School of Information Science and Technology, Harbin Institute of Technology, Shenzhen, China. Longyu Zhou is with Information System Technology and Design, Singapore University of Technology and Design and also with ChinaTelecom Singapore Innovation Research Institute. Tony Q. S. Quek is with Information System Technology and Design Pillar, Singapore University of Technology and Design, Singapore. (e-mail: liangth@hit.edu.cn; sunanchi@hit.edu.cn; pingyq@stu.hit.edu.cn; GuangyuLei@stu.hit.edu.cn; chenxinglin@stu.hit.edu.cn; zhoulyfuture@outlook.com; zhangtt@hit.edu.cn; zqy@hit.edu.cn; tonyquek@sutd.edu.sg.)
}

}

\markboth{IEEE Network}
{Shell \MakeLowercase{\textit{et al.}}: A Sample Article Using IEEEtran.cls for IEEE Journals}

\maketitle

\begin{abstract}
The emerging low-altitude economy has catalyzed the large-scale deployment of unmanned aerial vehicles (UAVs), driving a paradigm shift in environment monitoring, logistics, and emergency response. However, operating within these environments presents notable challenges as pervasive coverage holes, unpredictable interference, and spectrum scarcity. To this end, this article present a communication and control co-design framework to enable a resilient architecture for cellular-connected UAVs. Specifically, we first characterize typical service applications and their stringent performance requirements, followed by a comprehensive analysis of the unique challenges. To bridge the gap between volatile wireless links and rigid flight stability, a three layered architecture is proposed, integrating pre-flight strategic planning, in-flight adaptive action, and system-level resource orchestration. Furthermore, we detail the key enabling technologies for communication and control co-design. Preliminary case studies are proposed to validate that the co-design framework significantly improve the resilience of cellular-connected UAV systems, providing a robust foundation for the evolution of intelligent low-altitude networks.
\end{abstract}

\section{Introduction}
The emergence of the low-altitude economy has catalyzed a fundamental paradigm shift in the role of unmanned aerial vehicles (UAVs) from the height of 100 $\rm{m}$ to 3000 $\rm{m}$.  This evolution is driving various emerging applications such as urban air mobility, automated logistics, and collaborative emergency response, which accelerate the large-scale deployment of UAV in the low altitude airspace \cite{lthnetwork}.  To realize this vision, reliable communication and low latency control for are required for UAV operations. 
 
Obviously, ultra-reliable low-latency Communication (URLLC) is a good solution for enabling safe and scalable autonomous aerial systems. Inspired by the development of wireless communications in terms of bandwidth, capacity, coverage, and latency, the low-altitude wireless networks (LAWNs) are proposed recently to support the sensing, communication and control for the airspace UAVs \cite{yuan2025ground}. However, they are limited by wide deployment in a short time.  Consequently,  existing ground cellular networks (GCNs), with their wide-area coverage, security and reliability, high throughput, and ability in enhancing navigation accuracy, provide an important foundation for building LAWNs \cite{hdxnetwork}. Compared to dedicated or satellite networks,  GCNs for UAV control and communication can offer several decisive advantages.
\begin{itemize}
  \item {\bf Numerous Infrastructure and Cost-Efficiency}: Unlike dedicated UAV communication and control links requiring expensive point-to-point stations, GCNs leverage existing, globally standardized infrastructure, which can enable wide-area connectivity \cite{zywcm}.
  \item {\bf Low Latency for Real-time Control}: Satellite constellations, especially those in geostationary Earth orbit (GEO), suffer from inherent propagation delays exceeding 500 ms, which is incompatible with the millisecond-level requirements of URLLC. Even low Earth orbit (LEO) satellites face challenges with rapid handovers and high Doppler shifts. In contrast, GCNs offer low-latency air-ground interfaces essential for high-frequency control requirements.
  \item {\bf High Scalability and Interference Management}: Traditional private radio links operate on limited proprietary frequencies and struggle with mutual interference when multiple swarms occupy the same airspace. GCNs are built for massive multi-user access, employing advanced spatial multiplexing and sophisticated interference management methods \cite{fotosurvey}.
  \item {\bf Mobility Management and Edge Intelligence}: GCNs provide a mature framework for mobility management. Additionally, The integration of edge intelligence within cellular architectures allows for offloading heavy computational tasks from the resource limited UAV to the network edge .
\end{itemize}

Multiple research efforts, which can be divided to three directions, have explored the communication and control design for cellular-connected UAVs \cite{wyiot}. Control-aware strategies focus on advanced control laws, such as event-triggered control (ETC), model predictive control (MPC), and linear quadratic regulator (LQR), while communication impairments like latency, packet loss, and bandwidth limitations are treated as stochastic constraints. The objective is to maintain closed-loop stability and tracking precision despite the volatility of the cellular link. Communication-aware optimizations often prioritize the link performance, such as maximizing throughput and minimizing bit error rates (BER), while the control stability requirements are often simplified into minimum data-rate or maximum-delay thresholds \cite{lthiot}. Integrated communication and control optimization methods seek a fundamental trade-off between communication and control costs,  aiming to maximize the overall system utility, energy efficiency, and operational safety \cite{cyhwcl}. While effective in specific scenarios, they often rely on simplified assumptions and lack of consideration for the characteristics of low-altitude environments. Terrestrial base stations (BSs) in GCNs typically employ antennas with a down-tilt configuration to maximize ground coverage. Despite the high probability of line-of-sight (LoS) links at high altitudes, UAVs often find themselves in the blind spot of BS antenna patterns, leading to sudden signal dropouts and coverage holes in 3D space. Additionally, the good LoS conditions that improve signal strength also expose UAVs to signals from dozens of neighboring BSs. This results in severe interference, leading to the repeatedly handovers. Mostly importantly, the limited resource can not support the highly communication and control requirements all the time, especially in large-scale UAV swarms \cite{zscc}.  
\begin{figure*}[ht]
  \centering
  \includegraphics[width=1.9\columnwidth]{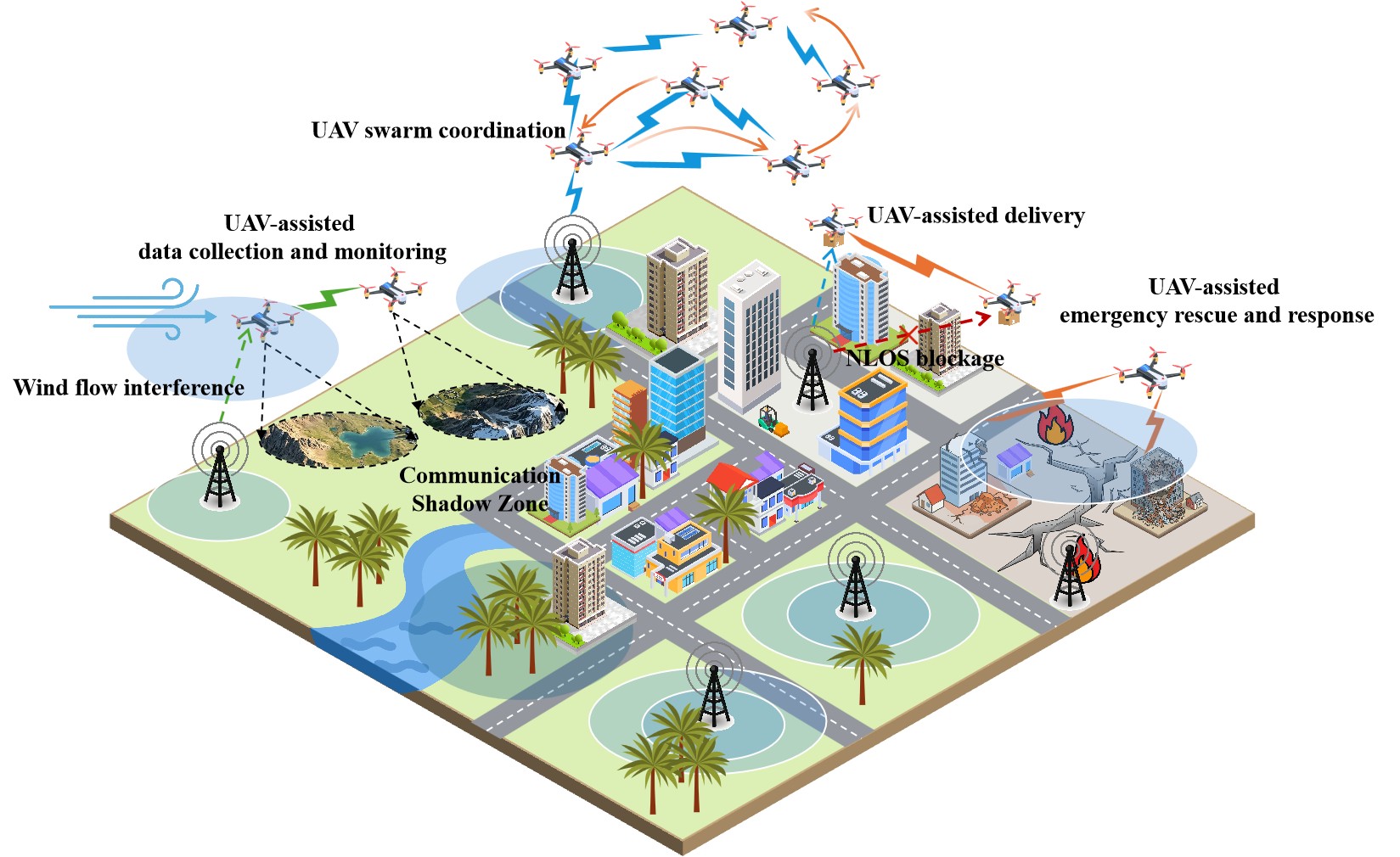}
  \caption{The illustration of emerging applications of cellular-connected UAVs in low-altitude economy.}\label{F1}
\end{figure*}
It should be noted that most existing solutions focus exclusively on a specific phase or a single layer of the UAV mission. While these investigations provide valuable theoretical insights, they fail to capture the dynamic interdependencies that evolve across the entire mission chain, which impacts their scalability and applicability. To fill this gap, this article proposes a holistic communication and control co-design architecture to fully excavate the potential {\it resilience} of cellular-connected UAV systems. This novel framework integrates the coordination of communication and control into a closed-loop task chain consisting of  pre-flight strategic planning, in-flight adaptive action, and system-level resource orchestration. It features key enabling technologies for communication and control co-design, including trajectory planning, adaptive control, and network scheduling. To have a comprehensive understanding, we present our two preliminary case studies of UAV trajectory tracking and UAV swarms control. 

The structure of the remainder of this article is as follows.  We first introduce the emerging applications and potential challenges for cellular-connected UAVs. Then, we elaborate the proposed the closed UAV communication and control co-design architecture. After that, the key enabling technologies are provided and analyzed. Following this, we give two case studies to showcase the results of co-design strategies. Finally, we draw conclusions and present some significant future works.

\section{Emerging Applications and Design Challenges}
Fig. \ref{F1} illustrates emerging applications and potential challenges for low-altitude cellular connected UAV. In these scenarios, the interplay between communication and control becomes the decisive factor for mission success. As summarized In Table \ref{tab:uav_comm_ctrl_requirements}, the communication and control requirements vary significantly across services, as defined in the 3GPP  technical report in Release 19 \cite{3gpp_tr_22_821,3gpp_ts_22_261}. In the following, we detail the advantages and challenges of those applications.
\begin{table*}
	\centering
	\caption{Summary of Communication and Control Requirements for Typical Applications in Low-Altitude UAV Networks}
	\label{tab:uav_comm_ctrl_requirements}
	\resizebox{\textwidth}{!}{%
		\begin{tabular}{@{}lccccccc@{}}
			\toprule
			\multirow{2}{*}{\textbf{Service Scenario}} & 
			\multicolumn{3}{c}{\textbf{Communication Requirements}} & 
			\multicolumn{2}{c}{\textbf{Control Requirements}} & 
			\multirow{2}{*}{\textbf{\makecell{Altitude \\ Requirement}}} & 
			\multirow{2}{*}{\textbf{\makecell{Speed \\ Requirement}}} \\
			\cmidrule(lr){2-4} \cmidrule(lr){5-6}
			& \textbf{\makecell{Data Rate \\ (UL/DL)}} & \textbf{\makecell{End-to-End \\ Latency}} & \textbf{\makecell{Communication \\ Reliability}} & \textbf{Control Accuracy} & \textbf{\makecell{Control Command \\ Latency}} & & \\
			\midrule
									\textbf{Emergency Rescue and Response} & 
			\makecell[c]{UL: $\ge$ 25 Mbps \\ DL: 300 Kbps} & 
			\makecell[c]{UL: 100 ms \\ DL: 20 ms} & 
			99.99\% & 
			\makecell[c]{Horizontal: 0.5 m \\ Vertical: 1 m} & 
			40 ms & 
			$<$ 300 m & 
			\makecell[c]{$\leq$ 160 km/h} \\
			\addlinespace

			\textbf{Data Collection and Monitoring} & 
			\makecell[c]{UL: 4 - 100 Mbps \\ DL: 600 Kbps} & 
			100 - 200 ms & 
			99.9\% & 
			\makecell[c]{Horizontal: 0.5 m \\ Vertical: 1 m} & 
			1 s & 
			$<$ 100 m & 
			$<$ 120 km/h \\
			\addlinespace
\textbf{Collaborative Swarm Coordination} & 
			\makecell[c]{UL/DL: 1 - 10 Mbps} & 
			$<$ 10 ms & 
			\makecell[c]{99.99\% - \\ 99.999\%} & 
			\makecell[c]{Horizontal: $<$ 0.1 m \\ Vertical: $<$ 0.1 m} & 
			\makecell[c]{$<$ 10 - 40 ms} & 
			30 - 300 m & 
			$<$ 60 km/h \\

			\addlinespace

						\textbf{Security Patrol} & 
			\makecell[c]{UL: 120 Mbps \\ DL: 50 Mbps - 300 Kbps} & 
			\makecell[c]{UL: 20 - 200 ms \\ DL: 20 ms} & 
			99.9\% & 
			\makecell[c]{Horizontal: 0.1 - 0.5 m \\ Vertical: 1 m} & 
			1 s & 
			30 - 300 m & 
			\makecell[c]{Avg. 60 km/h \\ Max. $<$ 120 km/h} \\
			\bottomrule
		\end{tabular}%
	}
\end{table*}

\subsection{Emergency Rescue and Response}
In emergency rescue missions, the UAVs typically with a high speed up to 160 $\rm{km/h}$ to reach the destination, requiring a command latency as low as 40 $\rm{m}$. These operations demand ultra-reliable command-and-control links to support millisecond-level maneuvering while simultaneously transmitting the high quality video data. The co-design of communication and control provides a significant advantage. For instance, by sharing real-time channel state information with the control layer, the UAV can actively adjust its flight trajectory to avoid the coverage blind zones and building occlusions shown in the diagram.

However, the primary challenge in emergency rescue lies in the unpredictable signal degradation caused by these physical barriers and blind coverage spots, where the absence of a direct LoS link can lead to transient yet catastrophic control failures. Additionally, maintaining the stringent requirements of  URLLC while managing high data transmission creates a competition for limited radio resources. 
\subsection{Data Collection and Monitoring}
UAVs are increasingly deployed for data collection and environmental monitoring. They need a horizontal control accuracy around 0.5 $\rm{m}$ and high uplink data rates about 100 $\rm{Mbps}$. These missions require a balance between data freshness, flight precision and energy consumption over extended durations. Such applications usually locate at remote areas with heavy weather interference or operate at altitudes lower than 100 $\rm{m}$ with severely signal block due to high buildings. Through a co-design strategy, the UAV can adaptively change its update frequency based on the current channel quality. This approach will significantly reduce the energy consumed by the communication and improve the control stability caused by adverse conditions.
 
The core design challenge here is the spatio-temporal trade-off between sampling frequency and energy efficiency. At altitudes below 100 $\rm{m}$, building occlusions and ground level interference are more serious. The co-design strategy must ensure that the control law remains robust against varying packet-loss rates, especially while the UAV operates at the edge of cellular coverage. This requires an adaptive framework capable of managing resources, ensuring that the high data transmission does not break the control stability.

\subsection{Swarm Coordination}
Large-scale swarm operations is the final vision of UAV operation, requiring high synchronization precision with latencies below 10 $\rm{ms}$ and reliability reaching 99.99\%. However, they are highly sensitive due to limited spectrum resources. A co-design strategy allows for control-aware resource orchestration, where the ground infrastructure prioritizes the communication resource for UAVs with higher control uncertainties. For example, when multiple UAVs congregate in a single cell, the network layer must intelligently allocate limited spectrum resource and design UAVs actions.

Nevertheless, the density of these swarms introduces the challenge of massive concurrent access and severe co-channel interference. As the swarm traverses ground cell boundaries, the high probability of LoS links at cruising altitudes (up to 300 $\rm{m}$) exposes the UAVs to signals from multiple neighboring BSs. This leads to severe handover issues, which cause link outages and over 10 $\rm{ms}$ latency budget.  Managing the association of UAVs while satisfying the finite-blocklength constraints of URLLC packets and specific control accuracy requires a high-level coordination strategy.

\section{A Closed-Loop Resilient Architecture}
\begin{figure*}[ht]
  \centering
  \includegraphics[width=2\columnwidth]{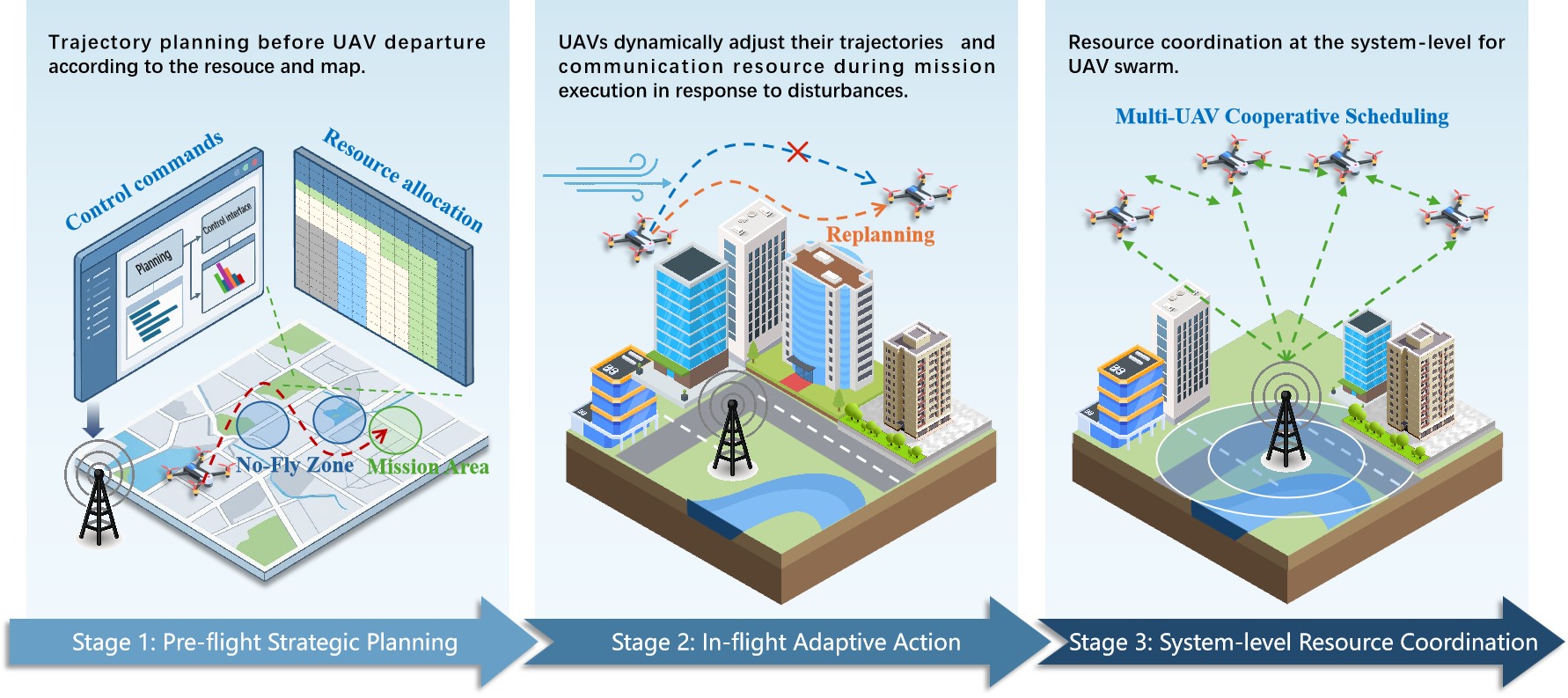}
  \caption{The illustration of our resilient communication and control framework for cellular-connected UAV systems.}\label{F2}
\end{figure*}
To effectively address the above challenges in low-altitude airspace, we propose a holistic, closed-loop resilient co-design architecture. As shown in Fig. \ref{F2}, we illustrate our framework with a rigorous temporal logic form, which consists of pre-flight strategic planning to in-flight tactical action, and finally to system-level resource orchestration. Our architecture ensures that the communication and control systems are not the independent processes but are deeply coupled during the entire mission execution, guaranteeing the resilience of UAV systems.

\subsection{Pre-flight Strategic Planning}
The first stage of the closed loop is the strategic planning before the UAV departure. Different with traditional methods with shortest distance, our design focuses on mitigating building occlusions and coverage blind zones based on the allocated resource and the priori knowledge including BSs deployment, radio map, and building distribution. 

In this stage, we treat the feasible cellular coverage radio map as a virtual physically region by treating 3D signal-interference-noise-ratio (SINR) distribution and its threshold as a fundamental navigation feasible region and constraint, respectively. This proactive approach ensures that the UAV is routed through high-quality links, effectively bypassing potential coverage holes and minimizing the frequency of unstable handovers, which provides the mission with a robust connectivity foundation.

\subsection{In-flight Adaptive Action}
Once the UAV flies in the airspace, the focus shifts to real-time resilience against the disturbance, such as sudden wind gusts and random obstacles. During this stage, the co-design of communication and control should adaptively balance the urgency of the control task with the time-varying wireless channel, such as trajectory replanning and communication resource re-allocation.

The core motivation here is to move away from rigid, periodic communication, which wastes UAV energy and communication resource.
Instead, we need a state-aware triggering logic to bridge the gap between co-design strategy and practical environment. In this situation, the UAV should adaptively adjust its communication frequency based on its actual control performance.  For example, when faced with weather disturbances, the UAV triggers high-frequency control updates to maintain tracking precision. This adaptation process ensures that the mission remains stable even when the network experiences the URLLC transient jitters or latency spikes characteristic.

\subsection{System-level Resource Orchestration}
The final stage of the closed loop operates at a global scale, addressing the challenges of limited spectrum resources and co-channel interference within large-scale swarm operations. This stage introduces a dynamic feedback loop between the network infrastructure and the collective state of all UAVs.

Rather than allocating bandwidth based solely on signal strength, the ground base station performs control-oriented resource orchestration. It monitors real-time control states and estimation errors of each UAV in the swarm. When a specific agent encounters severe interference or enters a high-precision maneuver, the network dynamically reallocates radio resources to that node. This system-level coordination prevents localized instabilities from cascading into a swarm failure, realizing a truly resilient and scalable low-altitude UAV systems.

\section{Key Enabling Technologies for Co-design}
The realization of resilient UAV systems requires a transition from traditional decoupled algorithms to cross-layer optimizations. This section details the key technologies of our framework and explores emerging enhancement methods.

\subsection{Connectivity based Trajectory Planning}
To address the pre-flight challenges, we can utilize graph-of-convex sets (GCS) as a powerful tool for global trajectory optimization. The 3D low-altitude airspace, cluttered with building occlusions and antenna nulls, is inherently non-convex and complex. GCS allows us to decompose this irregular space into a collection of convex regions where safe connectivity expressed by SINR over a certain threshold $\gamma_\text{th}$ is guaranteed. By formulating the pathfinding problem as a shortest-path search over these convex sets, we can leverage powerful semi-definite programming (SDP) solvers to find globally optimal trajectories that minimize both travel distance and connectivity risk. Unlike traditional sampling-based methods, GCS provides a deterministic guarantee that the UAV will remain within high-quality coverage zones, effectively integrating communication reliability into the flight path.

The decision in this stage is executed at the cloud, where the high computation requirements can be tolerated. Therefore, the resilience of this strategic planning process can be further augmented by the integration of digital twin (DT) and radio map technologies.  By offloading these high-complexity simulations to cloud servers, the system can construct a high-fidelity DT of the operational environment. The DT continuously updates the GCS model with real-time electromagnetic data, thereby reducing the probability of the UAV never enters a signal shadow unexpectedly.

\subsection{Adaptive Communication and Control Co-design}
After UAV taking off, the communication and control actions can be adjusted according to the sensing results of the UAV. In this stage, we turn the view on UAV itself. Therefore, the joint optimization  should carefully design to reduce the risk from uncertain environments. Specifically, to combat the stochastic nature of the cellular link, the UAV can employ packetized predictive control (PPC) strategy instead of transmitting a single control input $u_k$ at each time step. PPC acts as a temporal buffer by transmitting a sequence of future predicted control inputs as
\begin{equation}
U_k = \{u_{k|k}, u_{k+1|k}, \dots, u_{k+N|k}\}
\end{equation}
where $N$ is the predictive horizon. If a transient packet loss occurs due to unpredictable interference, the UAV can autonomously execute the previously received $u_{k+1|k}$ to maintaining stable flight. 

Furthermore, the advanced control method should be presented to reduce the unnecessary resource consumption and enhance the robustness of control. To optimize this process, we can employ the ETC, where the communication link is only activated if the discrepancy between the actual state and the predicted state exceeds a predefined threshold $\delta$. This method ensures that URLLC resources are used only when they are critically needed, preserving both energy and spectral efficiency. Moving beyond reactive triggers, self-triggered MPC (ST-MPC) determines the next trigger instant based on the UAV's internal dynamic model. This method predicts how long the current control sequence will remain valid before the state uncertainty becomes too high. This significantly reduces the need for continuous state monitoring, conserving both onboard sensing energy and network signaling.

This process can be further optimized via integrated sensing and communication (ISAC) technology \cite{lshiot}. By leveraging ISAC, the communication signal itself becomes a sensor. The system can utilize the echo of the transmitted waveform to detect real-time environmental changes, such as localized wind gusts or unmapped physical obstacles. This improved or additional sensing results are fed back into the ST-MPC or ETC mechanisms, significantly enhancing the resilience of UAV systems.

\subsection{Control-Oriented Scheduling for UAV Swarms}
During the operations of UAVs, there are other different devices accessing the same cell.  Therefore, the co-design strategy need to be moved from UAV-level to the network-level, especially in large-scale swarms. Different from tradition fairness-based scheduling, the resource allocation should be control-aware, which uses the metrics like control urgency of information (UoI), age of information (AoI) and value of information (VoI) \cite{zxtwc,lthtwc}.

In a dense swarm, the base station monitors the state estimation error variance $\Sigma_{i,k}$ for each UAV $i$. When a UAV experiences severe interference or enters a complex maneuver, its $\Sigma_{i,k}$ increases. The network then dynamically reallocates physical resource blocks and adjusts the modulation and coding scheme (MCS). Through this process, we can ensure that the limited spectrum is utilized to maximize the collective safety of the entire swarm, rather than just individual throughputs or control costs.
 
Furthermore, the intelligence behind this orchestration can be significantly augmented by multi-agent reinforcement learning (MARL) deployed at the network edge. By training on vast amounts of flight data, the MARL agent can predict which UAV is most likely to suffer a control failure. This predictive orchestration can prevent localized tracking errors from cascading into a swarm-level collision. This global feedback loop ensures that the finite spectrum resource is always directed toward maintaining the overall safety and resilience of the UAV system.

\begin{figure*}[t]
		\centering
		\includegraphics[width=0.85\textwidth]{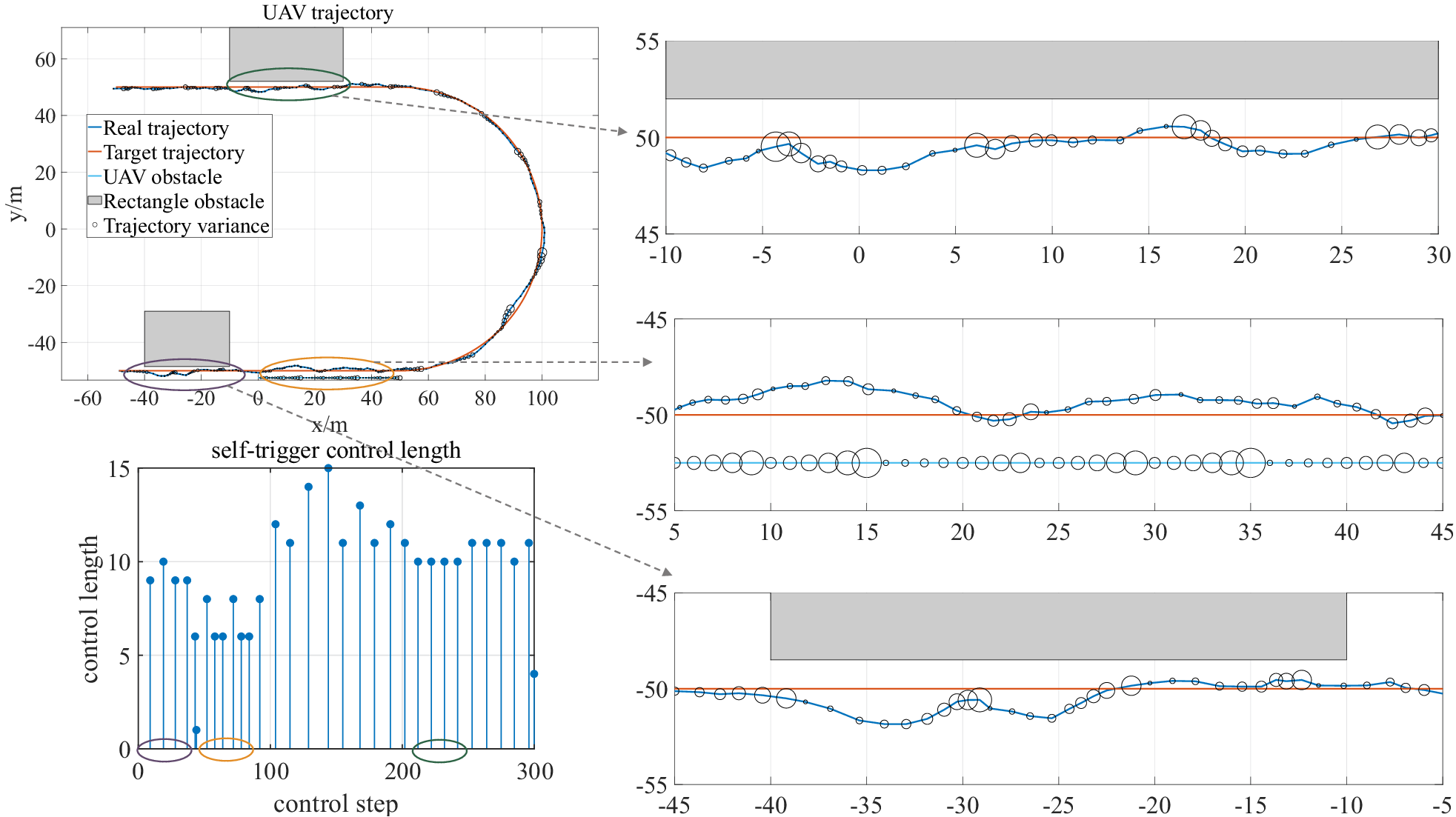}
		\caption{Adaptive communication and control co-design for a remote UAV system. The left panel shows the target and realized trajectories in the presence of static and dynamic obstacles. The lower-left panel shows the self-triggered control length over time. The three zoomed views on the right highlight representative operating regions.}
		\label{fig:adaptive_case_study}
	\end{figure*}
\section{Case Studies: From Single-UAV Co-Design to Multi-UAV Closed-Loop Scheduling}
This use case presents a transition from adaptive communication-control co-design for a single remotely controlled UAV to context-aware closed-loop scheduling for a UAV swarm. The first case focuses on how a controller can adapt the timing and length of control updates for one UAV operating in a dynamic environment. The second case considers a multi-UAV setting, where limited wireless resources must be shared across many UAVs and communication decisions become a scheduling problem at the swarm level.
	
\section*{Case I: Adaptive Communication and Control Co-Design for Remote UAV}

	Fig.~\ref{fig:adaptive_case_study} illustrates a remote UAV control framework in which a base station computes control actions and delivers them to the UAV through short-packet wireless transmissions. Instead of sending a new command at every slot, the controller transmits a segment of future control actions in one packet and adaptively decides how long that segment should be. Through this design, the system balances trajectory tracking, collision avoidance, and communication overhead in a unified manner, making remote UAV control more efficient under constrained wireless resources.
	
	As shown in Fig.~\ref{fig:adaptive_case_study}, the UAV is able to follow the target trajectory while avoiding both static rectangular obstacles and a moving UAV obstacle. The circles around the trajectory reflect how uncertainty evolves along the flight path, and the lower-left panel shows that the self-triggered control length changes continuously with the local control risk. In more challenging regions, where the UAV faces tighter safety margins or stronger uncertainty accumulation, the controller shortens the control interval so that updates arrive more frequently. In relatively benign regions, longer control segments are used to reduce communication burden while maintaining reliable tracking performance.
	
	The three zoomed views on the right further illustrate this adaptive behavior. Near obstacle-sensitive or high-risk segments, the controller behaves more conservatively to preserve safe margins for collision avoidance, whereas in safer segments it allows longer open-loop execution to improve communication efficiency. This case shows that communication and control should be designed jointly even for a single UAV, and that adaptive update lengths provide an effective way to support safe navigation with lower communication cost. In the reported results, this design reduces cumulative communication energy by {51.8\%} compared with conventional MPC.

\section*{Case II: Context-Aware Closed-Loop Scheduling for UAV Swarms}
\begin{figure*}[t]
	\centering
	\includegraphics[width=0.85\textwidth]{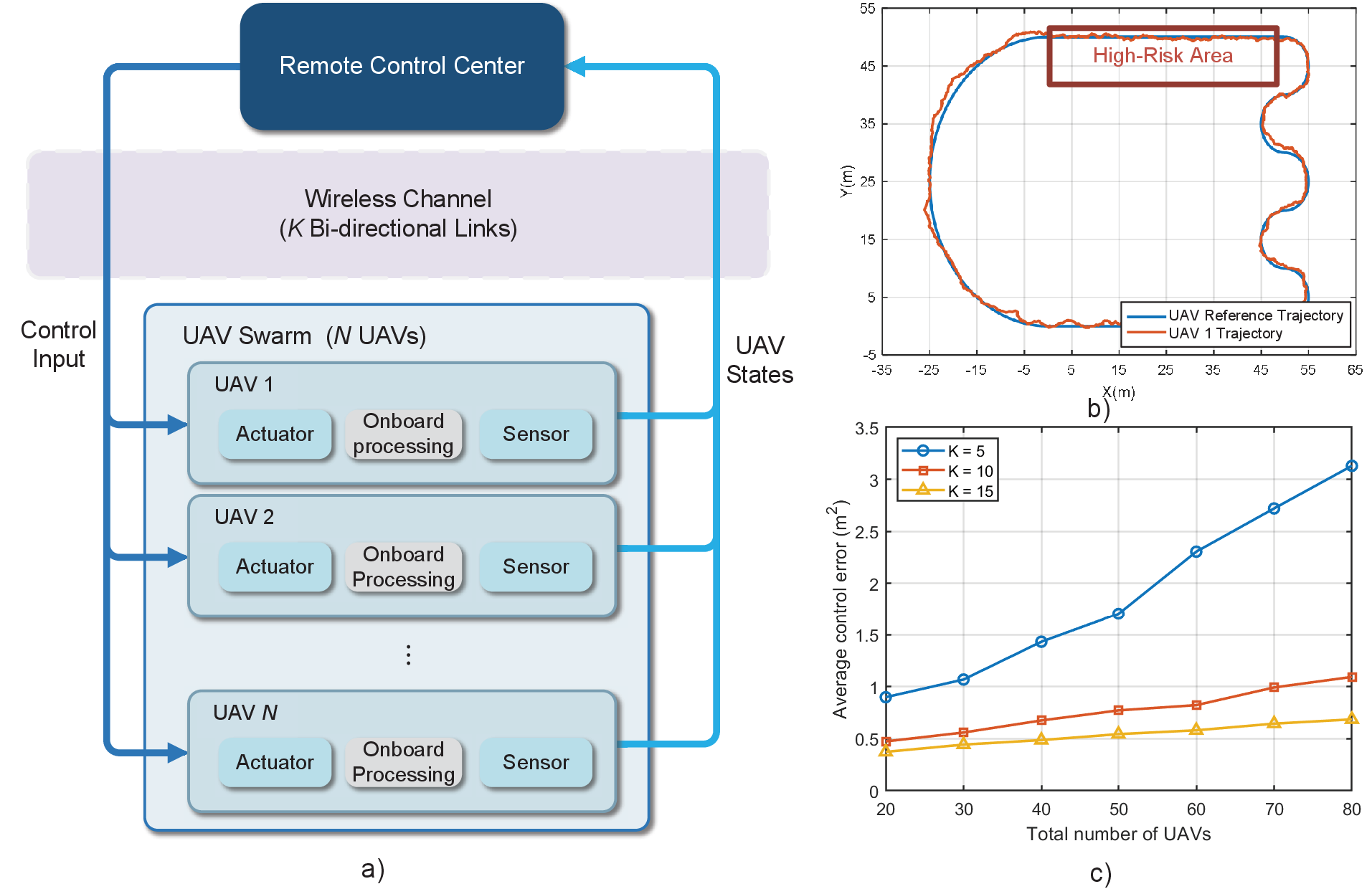}
	\caption{Case study of the proposed context-aware closed-loop scheduling framework for UAV swarms. (a) System architecture with a remote control node and a UAV swarm under limited bidirectional wireless links. (b) Trajectory tracking example, where the UAV follows the reference path and receives denser updates in the high-risk area. (c) Average control error versus the total number of UAVs under different numbers of UAVs that can access the channel in each time slot.}
	\label{fig:case_study}
\end{figure*}

This case study presents a context-aware closed-loop scheduling strategy for remotely controlled UAV swarms under limited wireless resources. As illustrated in Fig.~\ref{fig:case_study}(a), the system adopts a centralized architecture in which a remote control node coordinates a swarm of UAVs through limited bidirectional wireless links. In each time slot, only a subset of UAVs can upload their states and receive updated control commands, while the remaining UAVs continue operating based on predicted states and previously issued commands. Under this setting, the key challenge is to decide which UAVs should be prioritized for communication and control updates so that the overall swarm can still maintain reliable trajectory tracking.

To address this issue, the proposed design allocates communication resources according to the actual urgency of different UAVs rather than distributing them uniformly across the swarm. By jointly considering sensing freshness, control error, and environmental context, the framework gives higher priority to UAVs flying in more critical or demanding regions. Through the coordinated design of robust control and real-time update scheduling, the system is able to adapt communication decisions to flight conditions and control needs, making closed-loop swarm coordination more efficient under constrained wireless resources.

Fig.~\ref{fig:case_study}(b) and Fig.~\ref{fig:case_study}(c) highlight the performance of the proposed design under representative operating conditions. As shown in Fig.~\ref{fig:case_study}(b), the UAV is able to follow the reference trajectory with good accuracy, while denser updates are delivered in the high-risk area and other more demanding flight segments. Fig.~\ref{fig:case_study}(c) further shows that, as the total number of UAVs increases, the average control error gradually rises because more UAVs compete for the same communication resources, whereas allowing more UAVs to access the channel in each time slot can significantly reduce the error. These observations suggest that context-aware closed-loop scheduling provides an effective way to support reliable and scalable UAV swarm coordination under constrained wireless resources.
\section{Conclusions and Future Work}

In this article, we have presented a comprehensive framework of communication and control co-design for UAV systems. Specifically, we first provided the requirements of typical applications and analyzed the inherent challenges in the low-altitude airspaces. A resilient  architecture was present including strategic planning, tactical adaptation, and resource orchestration.
We further detailed the key enabling technologies that underpin this closed-loop framework. Two case studies demonstrated the effectiveness of proposed approach. 

Looking ahead, potential future works for the proposed architecture for are listed as follows.

\subsection{Joint Computing, Communication, Sensing, and Control Optimization}
The integration of computing, communication, sensing, and control should be investigated with a unified framework, where sensing provides environmental awareness, computing enables real-time processing, and communication ensures reliable data exchange. Developing a joint optimization method that balances the heterogeneous trade-offs among those different metrics will be a key research direction.
\subsection{Distributed Communication and Control Protocol}
As the increasing of UAV swarms, future work should focus on decentralized communication and control protocols  utilizing consensus-based algorithms. By developing distributed scheduling, UAVs can achieve collective coordination with reduced signaling overhead and increased tolerance to localized link failures. This direction will further enhance the resilience of UAV swarms.
\subsection{LLM-enabled Communication and Control}
The integration of large language models (LLMs) offers a transformative path for UAV operations \cite{yqwcm}. Future studies could explore how LLMs can serve as high-level decision maker that translate natural language mission goals into precise control tasks and communication strategies. 

\subsection{Integrated Space-Ground Network}
To eliminate the coverage blind zones and reach remote areas beyond terrestrial cellular footprints, the integration of space-air-ground integrated Networks (SAGIN) is useful. Utilizing LEO satellite constellations as non-terrestrial networks (NTN) backups can ensure ubiquitous connectivity for long-range UAV missions. The handover issue between GCN and NTN should be carefully investigated.


\bibliographystyle{IEEEtran}
\bibliography{ref}

\end{document}